\documentclass[aps,showpacs,prc,twocolumn,superscriptaddress,nobibnotes,nofootinbib]
{revtex4}
\usepackage[dvips]{graphicx}
\usepackage{epsfig}
\usepackage{bm}   % for adding bold symbols etc. in maths mode
\usepackage{dcolumn}% Align table columns on decimal point
\usepackage{graphicx}
\usepackage{rotate}
\usepackage{amsfonts}
\usepackage{amscd}

\begin{document}
\def\vlk{$V_{{\rm low}\,k}$}
\def\core{$^{88}$Sr}
\def\zr{$^{92}$Zr}
\def\nb{$^{93}$Nb}
\def\mo{$^{94}$Mo}
\def\ru{$^{96}$Ru}
\def\pd{$^{98}$Pd}
\def\cd{$^{100}$Cd}
\def\ms{2$^+_{1, {\rm ms}}$}
\def\s {2$^+_1$}
\def\veff{$V_{\rm eff}$}
\def\tp{$2^+_p$}
\def\tn{$2^+_n$}
\def\etal{\it et al.}
\def\pn{\it pn}
\def\prl{Phys.\ Rev.\ Lett.\ }
\def\np{Nucl.\ Phys.\ }
\def\prc{Phys.\ Rev.\ C }
\def\pl{Phys.\ Lett.\ }
\def\fm{fm$^{-1}$}

\title{Microscopic Restoration of Proton-Neutron Mixed
Symmetry \\ in Weakly Collective Nuclei}

\author{J.\ D.\ Holt}
\email{jholt@triumf.ca}
\affiliation{Department of Physics and Astronomy, SUNY, Stony Brook,
NY 11794-3800, USA}

\affiliation{TRIUMF, 4004 Wesbrook Mall, Vancouver, BC V6T 2A3,
Canada}

\author{N.\ Pietralla}
\affiliation{Department of Physics and Astronomy, SUNY, Stony Brook,
NY 11794-3800, USA}

\affiliation{Institut f\"ur Kernphysik, Technische Universit\"at
Darmstadt, 64289 Darmstadt, Germany}

\affiliation{Institut f\"ur Kernphysik, Universit\"at zu K\"oln,
50937 K\"oln, Germany}

\author{J.\ W.\ Holt}
\affiliation{Department of Physics and Astronomy, SUNY, Stony Brook, NY 11794-3800, USA}

\author{T.\ T.\ S.\ Kuo}
\affiliation{Department of Physics and Astronomy, SUNY, Stony Brook, NY 11794-3800, USA}

\author{G.\ Rainovski}
\affiliation{Department of Physics and Astronomy, SUNY, Stony Brook, NY 11794-3800, USA}

\date{\today}

\begin{abstract}
Starting from the microscopic low-momentum nucleon-nucleon interaction {\vlk}, we 
present the first systematic shell model study of magnetic moments and magnetic dipole transition 
strengths of the basic low-energy one-quadrupole phonon excitations in nearly-spherical nuclei.
Studying in particular the even-even
$N=52$ isotones from {\zr} to {\cd} , we find the predicted evolution of the predominantly 
proton-neutron non-symmetric state reveals a restoration of collective proton-neutron 
mixed-symmetry structure near mid-shell.  This provides the first explanation for the
existence of pronounced collective mixed-symmetry structures in weakly-collective nuclei.
\end{abstract}

\maketitle

Mesoscopic quantum systems such as Bose-Einstein condensates, superconductors,
and quark-gluon systems are some of the most intensely studied in contemporary
physics \cite{ketterle02,ginzburg04}. Their dynamical properties are
determined by the interplay and mutual balance of collective and
single-particle degrees of freedom. In two fluid systems such as atomic nuclei,
the presence of an isospin degree of freedom only serves to enhance this complexity. Of 
particular interest for understanding the physics of these systems is the microscopic origin of
those excitations possessing collective two-fluid character.
Collective quadrupole isovector excitations in the valence shell, so-called
mixed-symmetry states (MSSs) \cite{iachello84}, are the best-studied examples
of this class of excitations. A special type of MSS, the $1^+$ scissor mode,
was predicted to exist \cite{loiudice78} and discovered \cite{bohle} in atomic
nuclei. It is not surprising then, that analogous scissor-mode states have subsequently been 
found in other two-fluid quantum systems such as trapped Bose-Einstein condensed gases
\cite{bec1,bec2}, metallic clusters \cite{met}, and elliptical quantum dots
\cite{qdots}.

Even though MSSs are a common feature of two-fluid quantum systems, atomic
nuclei are still the primary laboratory in which our understanding of them can be shaped. 
In the interacting boson model of heavy nuclei (IBM-2) the definition of
MSSs is formalized by the bosonic $F$-spin symmetry \cite{iachello84}. It
arises predominantly from a collective coupling of proton and neutron
sub-systems, and when the proton/neutron (pn) valence spaces are large enough,
strong coupling can arise between them. Naturally then, the best examples of
pn-symmetric and MSSs would be expected at mid shells. However, pronounced
MSSs have recently also been observed in weakly collective nuclei especially
in $N=52$ isotones. Multiphonon structures of MSSs are observed in the
nucleus {\mo} \cite{pietralla99,pietralla00,fransen03}, in neighboring nuclei
\cite{pietrallaru,werner02,fransen05}, and recently the first MSSs in an odd-mass
nearly-spherical nucleus were identified in {\nb} \cite{orce06}.  
% These states are of particular interest because of their sensitivity to the valence shell 
%proton-neutron interaction.
Even though the experimental properties of MSSs in this region have been
well described within the framework of the IBM-2
\cite{iachellobook,vanisacker}, the nuclear shell model (SM)
\cite{werner02,lisetskiy}, and the quasiparticle-phonon model \cite{loiudice},
the question of how these states arise and evolve has not yet been answered.

In this Letter, we provide the first microscopic foundation for the
formation and evolution of the fundamental MSS of vibrational
nuclei, the one-quadrupole phonon {\ms} state, from SM calculations
using the low-momentum nucleon-nucleon (NN) interaction
{\vlk} \cite{bogner03}.  
%This cutoff-dependent interaction is derived from a  high-precision 
%NN interaction with the requirement that low-energy observables, such as the elastic scattering 
%phase shifts, be preserved below the momentum cutoff.
{\vlk} defines a new class of NN interaction with a variable momentum cutoff (or resolution scale)
that reproduces low-energy two-nucleon observables.
For $N=52$ isotones from the $Z=40$ to the $Z=50$ shell closures, we
investigate the structure of the two SM one-phonon 2$^+$ states 
(labeled $2^+_I$ and 2$^+_{II}$), 
where the $2^+_{II}$ state is connected to the $2^+_I$ state
by a strong $M1$ transition. The 2$^+_I$ and 2$^+_{II}$ SM states, of course, correspond to the 
experimentally-observed 
one-phonon symmetric {\s} and mixed symmetric {\ms} states, respectively. The evolution of these
states is seen by examining two of the most important observables:
$B(M1;2^+_{II} \rightarrow 2^+_I)$ values and magnetic moments. Our calculations
indicate that the quantity driving this evolution is the orbital proton contribution to the $M1$ 
transitions. Moreover, the main cause underlying the formation
of MSSs in these weakly collective nuclei is the approximate energy
degeneracy in the proton and the neutron one-phonon quadrupole
excitations. This mechanism can be considered as a {\it microscopic
restoration of proton-neutron symmetry}. Because of the universal
nature of MS states, these studies should serve as qualitative
predictions of MS formation and structural evolution in other
nuclear regions, and give parallel insight into analogous structures
in general mesoscopic quantum systems.

Our microscopic shell model calculations are based on the low-momentum NN interaction 
{\vlk}. Although various NN potentials have been developed which reproduce the NN data up to 
momenta of $\sim$ 2 {\fm} \cite{machleidt}, they differ in their treatment of the high-momentum 
modes, which are known to complicate many-body calculations.
%Even though various high-precision NN models have been developed that successfully 
%reproduce all free-space data up to the momentum scale of $\sim$ 2.1 {\fm} 
%\cite{machleidt}, they differ in how they treat the unconstrained high momentum details above 
%this point. 
Using the renormalization group, we start from one of the high-precision
interactions and integrate out the high-momentum components above a cutoff 
$\Lambda$ such that the physics below this cutoff is preserved.  It has been shown that 
as $\Lambda$ is lowered to 2.1 {\fm}, the {\vlk} interactions flow to a 
result largely independent of the input interaction \cite{bogner03,bogner03b}. 
Furthermore, {\vlk} is energy independent and thus suitable for SM calculations in any 
nuclear region \cite{bogner02}. For the present work we use a {\vlk} interaction derived from the 
% chiral N$^3$LO interaction using a variable cutoff.  In this manner we can test the cutoff
%dependence of the properties of MS states as discussed in \cite{holtnb}.
CD-Bonn \cite{cdbonn} potential
with a cutoff $\Lambda$=2.1 {\fm}. Using the two-body matrix elements of {\vlk}, we 
then derive a SM effective interaction {\veff} based on the folded diagram
methods detailed in \cite{kuo90}. Through this process we include the effects of core polarization 
to second order, which has been shown to be a reasonable approximation to all order effects in 
the absence of many-body interactions \cite{holtkbb}. Finally, we note that {\vlk} with a fixed 
cutoff has been successfully used in recent nuclear structure studies \cite{coraggio06}.

\begin{figure}
\begin{center}
\includegraphics[width=7.5cm,height=6cm]{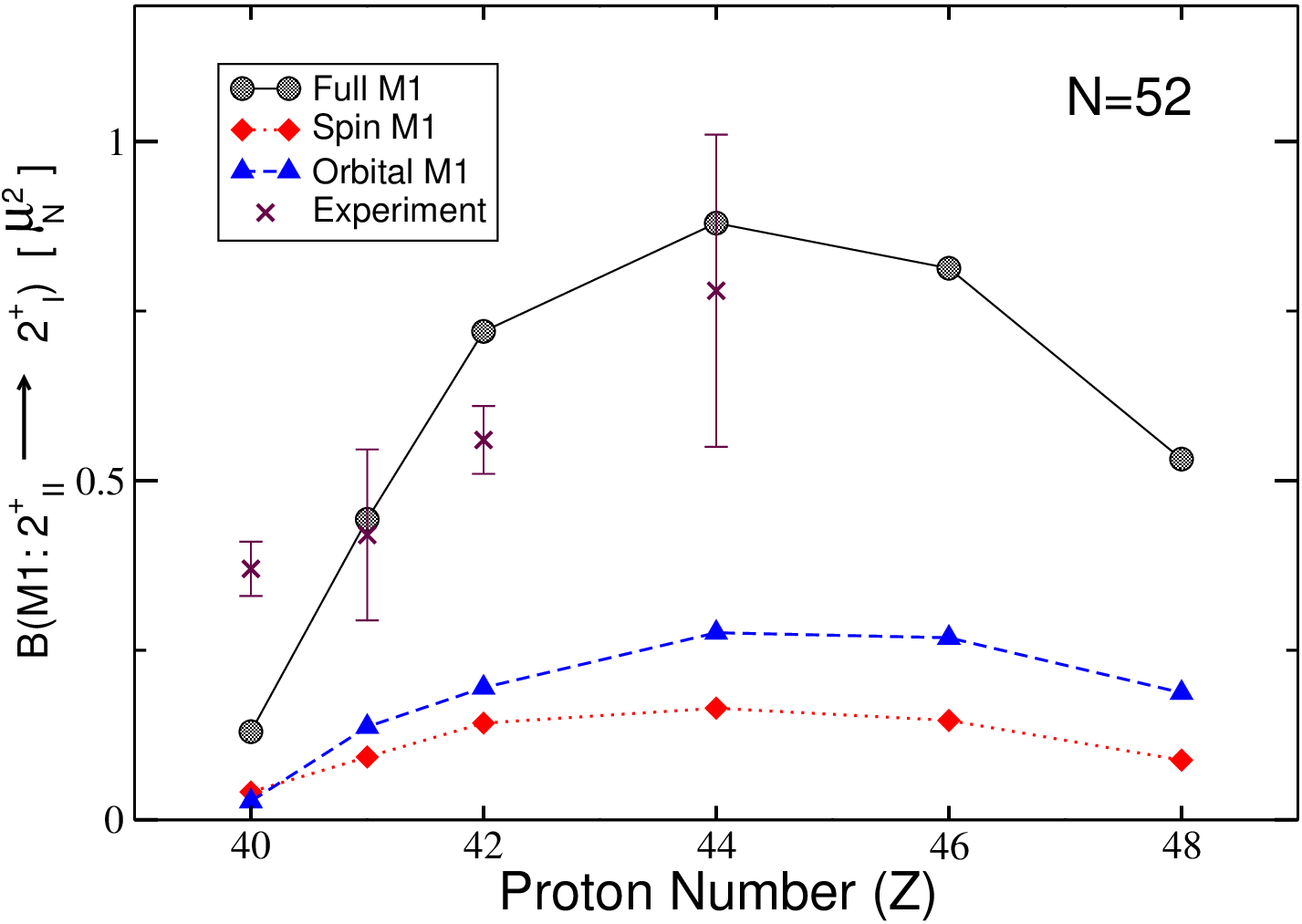}
\caption{(Color Online) Evolution of the total, orbital, and
spin $B(M1; 2^+_{II} \to 2^+_{I})$ values for $N=52$ isotones. The experimental values are from
\cite{fransen05,orce06,fransen03,pietrallaru}.}
\label{bm1}
\end{center}
\end{figure}

The present calculations were carried out with the Oxbash code \cite{oxbash} using
the effective interaction described above. {\core} was used as the inert core
and the proton-neutron model space taken to be: $\pi
:[2p_{1/2}$, $1g_{9/2}]$ and $\nu :[1g_{7/2}$, $2d_{5/2}$, $2d_{3/2}$,
$3s_{1/2}$, $1h_{11/2}]$, where the single particle energies were taken from the experimental 
values in $^{89}$Sr and $^{89}$Y.
%Values for the single particle energies of these orbits were taken from experiment.
%then optimized to best reproduce theexperimental spectra of $^{90}$Sr and $^{90}$Zr. 
Following this procedure we have obtained a good description of the extensive experimental 
data on MSSs for {\zr} and {\mo} \cite{holtnb} and predicted the properties of MSSs in the 
odd-mass nearly spherical nucleus {\nb} \cite{orce06,holtnb}.  It is debatable whether or not
the {\core} core is suitable for such calculations due to the close proximity of both the $2p_{3/2}$
and the $1f_{5/2}$ proton orbits.  We should emphasize, however, that our goal is not a highly 
accurate reproduction of experimental data, but to understand global features of MSSs in terms of
this simplified model space.  

The key experimental signature for MSSs is a strong $M1$ transition from the {\s} to the
{\ms} state, due to the isovector character of the $M1$ operator. The strength
and fragmentation of this transition is an indicator of the purity of the
MSSs. Bare orbital gyromagnetic factors $g^l_{\pi}=1\mu_N$, $g^l_{\nu}=0$, with
empirical (but not fine-tuned to fit the data) spin factors $g^s_{\pi}=3.18\mu_N$, 
and $g^s_{\nu}=-2.18\mu_N$ were used for calculating our SM
$B(M1;2^+_{II}\rightarrow2^+_I)$ values shown in \mbox{Fig.\ \ref{bm1}}.
They show a pronounced parabolic behavior, maximized at mid-subshell. Also
plotted in \mbox{Fig.\ \ref{bm1}} are the spin and orbital contributions,
both exhibiting an approximately parabolic shape.

The increase in $M1$ strength from {\zr} to {\ru} is in good qualitative
agreement with the data, though the SM calculations underpredict the
value in {\zr} and somewhat overpredict it in {\mo}. In \mbox{Fig.\
\ref{bm1}}, for comparison purposes, we have also included the average value for the 
$M1$ transitions from the mixed-symmetry doublet states, $\frac{5}{2}^-_{\rm ms}$ 
$\frac{3}{2}^-_{\rm ms}$, to the like-$J$ one-phonon states in {\nb} \cite{orce06};
it is clearly in general agreement with the overall trend. For
{\pd} and {\cd}, however, a decrease in $M1$ strength is predicted.

\begin{figure}
\includegraphics[height=6cm, width=7.5cm]{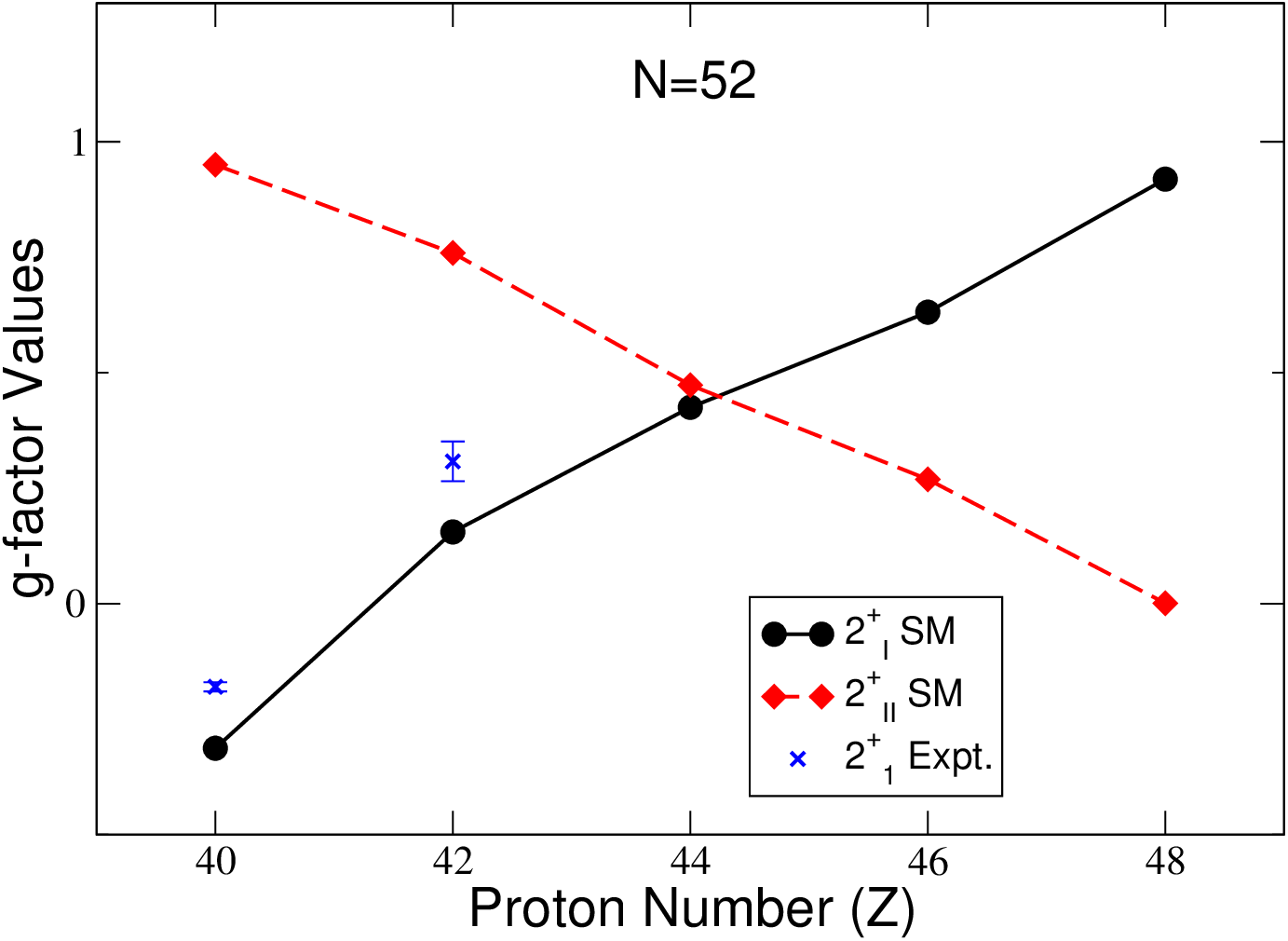}
\caption{(Color Online) Predicted evolution of the $g$-factors for
the $2^+_I$ and $2^+_{II}$ states across the series of N=52 isotones. The experimental values
are from \cite{jakob,mantica}.}
\label{gfactors}
\end{figure}

Figure \ref{gfactors} shows the predicted evolution of the $g$-factors of the
$2^+_I$ and $2^+_{II}$ states. As proton number increases the $2^+_I$
$g$-factors increase almost linearly while the $2^+_{II}$ $g$-factors decrease
linearly with approximately the same absolute slope. As a result they cross at
mid-shell. The calculated $g$-factors in {\ru} for the $2^+_I$ and
$2^+_{II}$ states are close (0.42 and 0.48, respectively) and not far from
the $Z/A=0.458$ value expected for a fully collective state. The negative
$g$-factor (-0.31) predicted for the $2^+_I$ state in {\zr} is in qualitative
agreement with experiment \cite{werner02,jakob}, indicating significant
neutron character of this state. At $Z=40$ the $2^+_I$ state of {\zr} is
expected to be primarily neutron in character. In drastic
contrast, we predict a large positive $g$-factor for the $2^+_{II}$
state in {\zr}, which suggests a dominant proton character,
from the $J=2$ coupling of protons in the $\pi (g^2_{9/2})$
orbit. The unbalanced proton-neutron content of the {\zr} $2^+_I$ and
$2^+_{II}$ states implies severe breaking of $F$-spin symmetry, not a
surprise for such a weakly collective system. This $F$-spin breaking is
reflected in the small $M1$ transition strength (see \mbox{Fig.\ \ref{bm1}}).
The changes in the $M1$ strengths and the $g$-factors for the $2^+_I$ and
$2^+_{II}$ states from {\zr} to {\cd} isotopes consistently show the
evolution of pn-symmetry character of the states: the purity of the MS character of the
$2^+_{II}$ state peaks in the mid-shell region before waning at the approach
of the Z=50 shell closure.

\begin{figure}
\includegraphics[height=6cm, width=7.5cm]{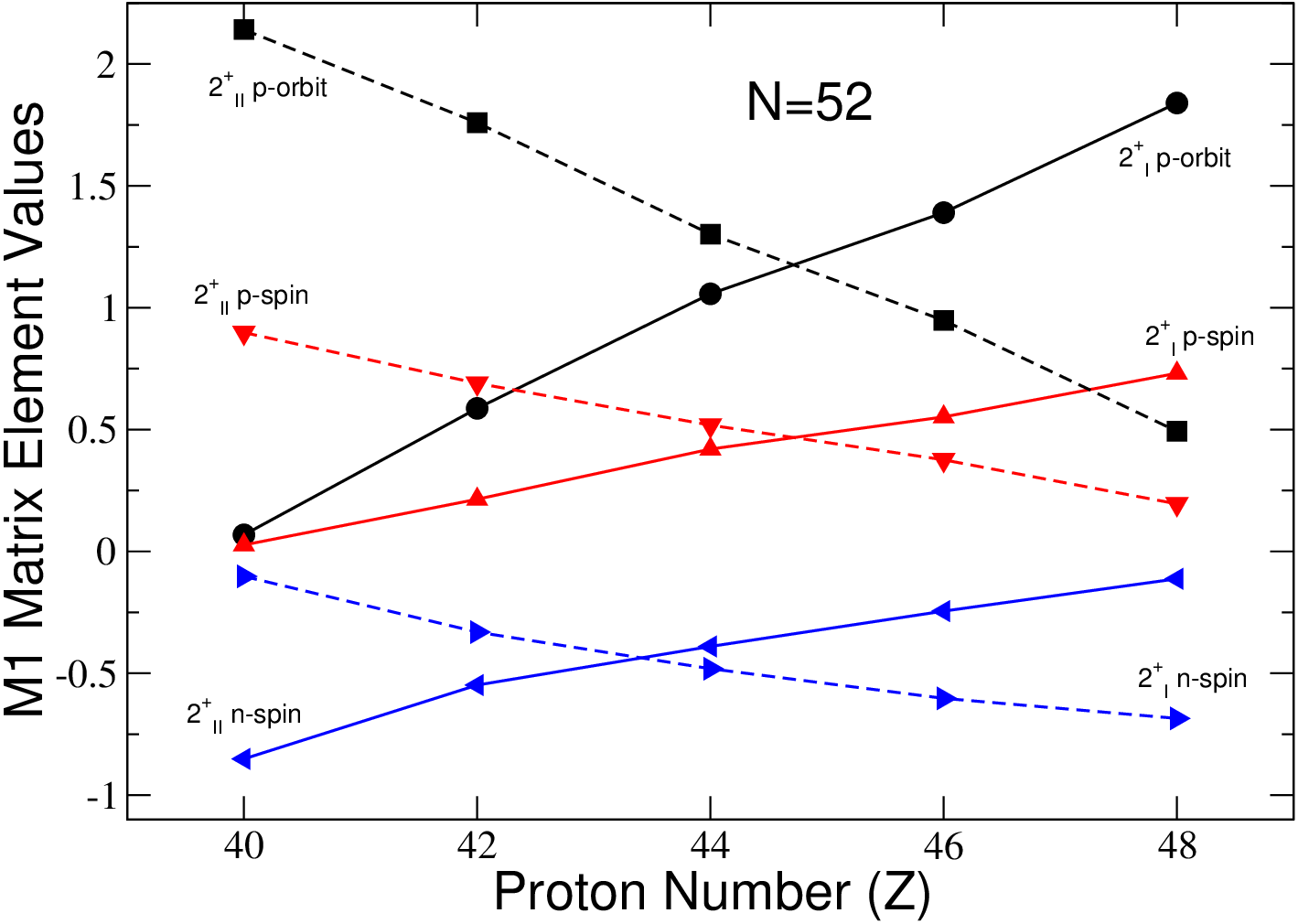}
\caption{(Color Online) Calculated proton/neutron--orbital/spin matrix elements contributing to the
 $g$-factors.}
\label{pnme}
\end{figure}

The p/n--orbit/spin contributions to the $M1$ matrix elements,
\mbox{$\langle
2^+_i\|g^{l,s}_{\pi,\nu}\cdot(M1)^{l,s}_{\pi,\nu}\|2^+_i\rangle_{(i=I,II)}$}
are shown in \mbox{Fig.\ \ref{pnme}}.
Considering first the $2^+_I$ states, we see
that in {\zr}, the neutron-spin ($g^s_{\nu} \cdot M1^s_{\nu}$)
component dominates, resulting in the negative $g$-factor. In {\mo},
however, the $M1^l_{\pi}$ contribution has risen while the
$M1^s_{\nu}$ has declined, yielding a small and slightly
positive $g$-factor. For the rest of the isotones, the $M1^l_{\pi}$
component continues increasing, while the others become more
negligible. For the  $2^+_{II}$ states, the opposite situation is seen.
The $M1^l_{\pi}$ term remains the driving force behind the
evolution of the $g$-factor, starting at a large, positive value in
{\zr}. It remains dominant until near $Z=50$ where it has decreased
to near zero while the $M1^s_{\nu}$ element has become negative enough to influence
the $g$-factor. The observed prominence of the orbital matrix element further confirms the
collective nature of these excitations.  

The calculated SM wavefunctions are generally complicated, but as
discussed in \cite{lisetskiy}, MS structure is evident in the
dominant components. Table \ref{mswf} gives the dominant amplitudes
for the normalized $2^+_I$ and $2^+_{II}$ wavefunctions. With the
exception of the $2^+_I$ state in {\zr}, the significant
contributions come from the
$\pi(p^2_{1/2}g^k_{9/2})_{0}\nu(d^2_{5/2})_{2}$ and
$\pi(p^2_{1/2}g^k_{9/2})_{2}\nu(d^2_{5/2})_{0}$ configurations,
where $k=2$ for {\mo}, $k=8$ for {\cd}, and the protons are jointly
coupled to either $J=0$ or $J=2$. For {\zr} we have only listed the
$\pi(g^2_{9/2})\nu(d^2_{5/2})$ configuration since the two protons
in $\pi(p^2_{1/2})$ cannot couple to $J=2$. These components only
represent a part of the full wavefunctions, but the percentage is
typically on the order of 60\% to 70\%. From \mbox{Table \ref{mswf}}
we see that the dominant parts of the $2^+_I$ and $2^+_{II}$
wavefunctions generally exhibit a pn-symmetric structure. Even with
these drastically truncated wavefunctions, the $2^+_I$ and
$2^+_{II}$ are formed by the same dominant components and are 
approximately orthogonal: we find
$\langle2^+_I|2^+_{II}\rangle<0.1$ in every case. The evolution of
the $M1$ strength and the $g$-factors is apparent -- at the low end
of the isotone chain, the $2^+_I$ state is predominantly neutron in
character and the $2^+_{II}$ state is largely proton. In {\mo}
already both components are more equally important. For the $2^+_I$
state, there is a linear increase in proton character and decrease
in neutron character across the shell with the opposite behavior
seen in the $2^+_{II}$ states. At mid-shell in {\ru} both states
have almost equal absolute amplitude of proton and neutron
excitations forming the fully symmetric and mixed symmetry 
one-quadrupole phonon states.

\begin{table}
\caption{Amplitudes of the dominant SM configurations contributing to the
$J^\pi = 2^+$ one-phonon wavefunctions.}
\vskip.1in
\label{mswf}
\begin{tabular}{cccc}
\hline \hline

Nucleus    &  Wavefunction Components   &  2$^+_I$ & 2$^+_{II}$  \\
[.05in]
%$^{92}$Zr  &  $ \pi \left(p^2_{1/2}\right)_0 \, \nu\left(d^2_{5/2}\right)_2$
%           &  0.815 & 0.286 \\[.1in]
{\zr}      &  $ \pi \left(g^2_{9/2}\right)_0 \, \nu\left(d^2_{5/2}\right)_2$
           &  0.462 & -0.129 \\[.1in]
           &  $ \pi \left(g^2_{9/2}\right)_2 \, \nu\left(d^2_{5/2}\right)_0$
           &  0.078 & 0.725 \\[.1in]

$^{94}$Mo  & $ \pi \left(p^2_{1/2}g^2_{9/2}\right)_0 \, \nu\left(d^2_{5/2}\right)_2$
           &  0.682 & -0.461 \\[.1in]
           & $ \pi \left(p^2_{1/2}g^2_{9/2}\right)_2 \, \nu\left(d^2_{5/2}\right)_0$
           &  0.426 & 0.652 \\[.1in]
$^{96}$Ru  & $ \pi \left(p^2_{1/2}g^4_{9/2}\right)_0 \, \nu\left(d^2_{5/2}\right)_2$
           &  0.586 & -0.584 \\[.1in]
           & $ \pi \left(p^2_{1/2}g^4_{9/2}\right)_2 \, \nu\left(d^2_{5/2}\right)_0$
           &  0.512 & 0.548 \\[.1in]
$^{98}$Pd  & $ \pi \left(p^2_{1/2}g^6_{9/2}\right)_0 \, \nu\left(d^2_{5/2}\right)_2$
           &  0.510 & -0.681 \\[.1in]
           & $ \pi \left(p^2_{1/2}g^6_{9/2}\right)_2 \, \nu\left(d^2_{5/2}\right)_0$
           &  0.576 & 0.448 \\[.15in]
$^{100}$Cd & $ \pi \left(p^2_{1/2}g^8_{9/2}\right)_0 \, \nu\left(d^2_{5/2}\right)_2$
           &  0.376 & -0.787 \\[.1in]
           & $ \pi \left(p^2_{1/2}g^8_{9/2}\right)_2 \, \nu\left(d^2_{5/2}\right)_0$
           & 0.638 & 0.305 \\[.1in]

\hline
\hline
\end{tabular}
\end{table}

These approximated wavefunctions and their $M1$ matrix elements can be
considered in the seniority scheme. The proton and neutron contributions to
the predominant seniority-two ($\nu = 2$) parts of the one-quadrupole phonon
wavefunctions in this sequence of isotones can be related to the fractional
filling $f = (Z-40)/10$ of the $\pi(g_{9/2})$ orbital with protons. The
wavefunctions can be approximated as
\begin{eqnarray}
\label{wf}
|2^+_I\rangle & \approx &
  \sqrt{f} |2^+_{\pi}\rangle + \sqrt{1-f} |2^+_{\nu}\rangle \\
\nonumber
|2^+_{II}\rangle  & \approx &
  \sqrt{1-f} |2^+_{\pi}\rangle - \sqrt{f} |2^+_{\nu}\rangle,
\end{eqnarray}
where the $|2^+_{\pi(\nu)}\rangle$ represents a SM configuration in which the
protons (neutrons) are coupled to $2^+$ and the neutrons (protons) are
coupled to $0^+$. To the extent that this two-state decomposition represents
a large percentage of the total wavefunctions, we can deduce
\begin{eqnarray}
\label{eq:g1}
g(2^+_I)         & = &
   \sqrt{\frac{2\pi}{45}} \left[\mu_\nu + f (\mu_\pi - \mu_\nu)\right] \\
\label{eq:gm}
g(2^+_{II}) & = &
   \sqrt{\frac{2\pi}{45}} \left[\mu_\pi - f (\mu_\pi - \mu_\nu)\right] \\
\label{eq:BM1}
B(M1;2^+_{II}\rightarrow 2^+_I) & = &
   \frac{1}{5} f(1-f) (\mu_\pi - \mu_\nu)^2
\end{eqnarray}
with $\mu_{\pi(\nu)} = \langle 2^+_{\pi(\nu)} \parallel M1 \parallel
2^+_{\pi(\nu)} \rangle$ being the diagonal $M1$ matrix elements of the
$\nu = 2$ proton (neutron) $2^+$ configuration. Note that $\mu_{\nu} < 0$ is
a constant for the $N=52$ isotones and, within the seniority scheme, the
matrix element $\mu_{\pi} > 0$ is independent of the filling of the proton
orbital and, hence, a constant as well. It immediately follows that the
$g$-factor of the $2^+_I$ state increases linearly with the filling of the
$\pi(g_{9/2})$ orbital while the $g$-factor of the $2^+_{II}$ state linearly
drops, instead, with the same absolute slope. In contrast to this, the $M1$
transition strength is proportional to the factor $f(1-f)$ and exhibits a
parabolic collective behavior over the $\pi(g_{9/2})$ shell with a maximum at
mid-shell. The size of the $B(M1;2^+_{II}\rightarrow 2^+_I)$ value is
proportional to the quadratic slope of the $g$-factors. This relation is
quantitatively ($\mu_\pi - \mu_\nu \approx 4.2\,\mu_N$) in good agreement
even with the full calculations.

Finally, the shell model results are qualitatively at variance with
the $F$-spin limit of the IBM-2. With effective boson $g$-factors
$g_\rho \equiv \sqrt{2 \pi/45}\,\mu_\rho$ the predictions of the
$F$-spin limit of the U(5) dynamical symmetry limit are
\begin{eqnarray*}
g(2^+_1)         & = &
   \sqrt{\frac{2\pi}{45}} \left[\mu_\nu + \frac{N_\pi}{N}
    (\mu_\pi - \mu_\nu)\right] \\
g(2^+_{1\rm,ms}) & = &
   \sqrt{\frac{2\pi}{45}} \left[\mu_\pi - \frac{N_\pi}{N}
    (\mu_\pi - \mu_\nu)\right] \\
B(M1;2^+_{1\rm,ms}\rightarrow 2^+_1) & = &
   \frac{1}{5} \frac{N_\pi}{N}(1-\frac{N_\pi}{N}) (\mu_\pi - \mu_\nu)^2
\end{eqnarray*}
which is formally identical to
Eqs.~(\ref{eq:g1}--\ref{eq:BM1}) with the replacement $f \rightarrow
N_\pi/N$ and $2^+_I=2^+_1$, $2^+_{II}=2^+_{1,\rm ms}$. However, due to
the convention of counting bosons as half of the number of
valence particles or holes outside the nearest closed shell, the boson
number fraction $N_\pi/N$ does not correspond to the fractional
filling $f$ of the proton shell but is instead for $N_\nu = 1$
approximately equal to the fractional half-filling. The $F$-spin limit
of the IBM-2 together with the conventional counting of bosons would
lead to local extrema for the $g$-factors of the one-phonon $2^+$
states at mid-shell [maximum for $g(2^+_1)$ and minimum for $g(2^+_{1,
{\rm ms}})$] and to a reduction of the $B(M1;2^+_{1\rm,ms}\rightarrow
2^+_1)$ value from $^{94}$Mo to $^{96}$Ru in contradiction to the
data.

The discrepancy in predicted evolution between the $F$-spin limit of
the IBM-2 and the SM calculation using {\vlk}
originates from the breaking of $F$-spin symmetry and
from its specific restoration near mid-shell in the shell model. The
smooth change in the proton-neutron character of the $2^+_I$ and
$2^+_{II}$ states, evident from \mbox{Table \ref{mswf}}, can be
attributed to the variation of energies of the one-phonon proton and
neutron configurations with filling of the $\pi(g_{9/2})$ orbital.
This can be seen in the data for the semi-closed shell nuclei in the
vicinity of the $N=52$ isotones. The $2^+_1$ energy of the even
$N=50$ isotones is indicative of the energy of the $2^+_\pi$
excitation. It drops slightly from 1509 keV in $^{92}$Mo to 1394 keV
in $^{98}$Cd. On the other hand, the energies of $2^+_1$ states in
the $^{92}$Zr and $^{102}$Sn are indicative of the energy of the
$2^+_\nu$ excitation. The data increase from 934 keV in $^{92}$Zr to
1472 keV in $^{102}$Sn. The difference in energy of these basic
$2^+_{\pi(\nu)}$ excitations translates into the IBM-2 framework in
a difference of proton and neutron $d$-boson energies that cause a
breaking of $F$-spin symmetry. Due to the low collectivity near
shell closure, this breaking cannot be restored by strong enough
pn-coupling. As a result, $F$-spin breaking is most pronounced in
$^{92}$Zr \cite{fransen05} and must be expected for $^{100}$Cd, too.
The energy crossing of the $2^+_\pi$ and $2^+_\nu$ configurations
near mid-shell, where the collectivity also peaks (see \mbox{Table
\ref{mswf}}), causes a specific restoration of $F$-spin symmetry,
yielding there the largest $B(M1)$ values and almost equal 
$g$-factors. This process is microscopic in origin and differs from
the usual $F$-spin symmetry generation due to the collective
pn-coupling in the framework of the IBM-2.

In summary we have used the low-momentum NN interaction {\vlk}
to provide a microscopic
description of the evolution of MSSs and magnetic dipole collectivity -- 
studying the $N=52$ isotones as a particular example.
The predicted observables reveal a new specific restoration
of proton-neutron symmetry which originates from the
energy degeneracy of basic proton and neutron excitations.
This process offers for the first time an explanation for the existence
of pronounced {\ms} structures in weakly collective nuclei and
might be observable in other two-fluid quantum systems.

\begin{acknowledgments}
We would like to thank B.\ A.\ Brown for his invaluable assistance with
the Oxbash code.  Support from the US Department of Energy under contract
DE-FG02-88ER40388 and the Natural Sciences and Engineering Research Council of Canada
(NSERC) 	is gratefully acknowledged.  TRIUMF receives federal funding via a contribution 
agreement through the National Research Council of Canada.
\end{acknowledgments}

\end{document}